\documentclass{ceurart}
\usepackage{booktabs}
\usepackage{caption}
\usepackage{graphicx}
\usepackage{natbib}
\usepackage{libertinus}

\usepackage{amsmath}
\usepackage{subcaption}
\begin{document}
	
\copyrightyear{2025}
\copyrightclause{Copyright for this paper by its authors. Use permitted under Creative Commons License Attribution 4.0 International (CC BY 4.0).}
\conference{LLAIS 2025: Workshop on Large Language Model Agents for Intelligent Systems, October 25, 2025, Bologna, Italy}
	
\title{Social Behavior Among Autonomous AI: How Large Language Models Interact in Dynamic Networks}
	
\author[1]{Narges Fardnia}[%
email=narges.fardnia@gisma-student.com
]
\author[2]{Fatemeh Seyedin}[%
email=fatemeh.s.seyedin@gmail.com
]
\author[2]{Matthias Becker}[%
email=xmb@hci.uni-hannover.de
]
\author[1]{Mahmoudreza Babaei}[%
email=mahmoudreza.babaei@gisma.com
]
\author[3]{Adrian Weller}[%
email=aw665@cam.ac.uk
]

\address[1]{Gisma University of Applied Sciences, BRAIN Research Center, Potsdam, Germany} 
\address[2]{Leibniz University Hannover, Germany}          
\address[3]{University of Cambridge, Cambridge, United Kingdom}

\begin{abstract}
Cooperation is a cornerstone of human societies, enabling collective progress in dynamic and uncertain environments. With the advent of AI systems acting autonomously, it becomes crucial to understand not only human-AI cooperation but also AI-AI interactions in adaptive networks.
In this work, we examine the interactions of AI using Large Language Models—Mistral, Llama3, Gemma3, and Phi3—in a public goods game within dynamic network structures. Our experiments were conducted under single-model and mixed-model conditions across Watts-Strogatz (WS), Barabási-Albert (BA), and Erdős-Rényi (ER) networks. We analyzed the impact of model architecture, network topology, and prompt design on cooperative behavior. Results show that Mistral and Llama3 offer high cooperation rates, while Phi3 shows defective tendencies. Additionally,  the random structure of Erdős-Rényi networks dramatically improves cooperation. Prompt design also plays a key role; a society-benefits prompt leads to a higher cooperation level. These findings offer a preliminary framework for LLM-based simulations in adaptive social networks.
\end{abstract}
	
\begin{keywords}
AI \& AI Interaction, Large Language Models \sep Public Goods Game \sep Dynamic Networks 
\end{keywords}
	
\maketitle
\section{Introduction}

Imagine walking into your office one day and finding out your new teammates are robots. Would you immediately feel comfortable handling an important project for them? Or would a small voice inside you wonder if you could really count on them? It might sound like science fiction, but it's quickly becoming our reality. AI is no longer something out of a movie—it is in our cars, guiding them through traffic; it is in our homes and pockets, answering questions, and helping us get things done. Tools like GPT are becoming everyday companions. As machines take on more roles alongside us, the big question is no longer if we will work with them, but how. This shift is changing not just how we work, but how we live, connect, and shape the world around us.

Teamwork has always been the secret sauce of human progress \cite{harari2011sapiens}. Think about it: from the first tiny molecules working together to form cells, to the massive cities we have built through countless people collaborating, cooperation fuels innovation and helps us specialize. But working together is not always easy. Take the classic "prisoner's dilemma"—two suspects, unable to talk, have to decide whether to cooperate or backstab each other. In a one-turn scenario, self-interest often outweighs cooperation, so the rational choice is to defect—even if teamwork would lead to better outcomes for everyone.
However, humans usually start off wanting to cooperate, maybe because of a twinge of guilt or a desire to help others, but if trust starts to erode, defection can quickly follow. As we increasingly interact with AI, a new wrinkle appears: we often trust machines less, are quicker to break promises to them, and tend to see them as less intelligent or cooperative than our human counterparts.
But it is not all skepticism! Sometimes people are totally on board with machine collaboration. Research, like that from Bonnefon and Rahwan ~\cite{rahwan2019machine, awad2018moral}, even suggests that humans and machines can work together beautifully with the right algorithms. In dynamic networks where individuals choose their partners, those who are skeptical of machines may opt for human-only groups. Others, though, might welcome AI into their teams, potentially creating more cooperative worlds. This brings up a fascinating question: How do AI systems cooperate with humans? Beyond just humans and AI, we also need to understand how different AI systems interact with each other when humans increasingly rely on them to get things done.

Our study dives into this question by observing how Large Language Models (LLMs)—specifically Mistral, Llama3, Gemma3, and Phi3 (free models)—behave in a "public goods game" within dynamic network structures. We focus on these models because they are freely available, with plans to investigate more widely used models in future work.
We set up experiments with either a single AI model or a mix of them across different network types: Watts-Strogatz, Barabási-Albert, and Erdős-Rényi. We wanted to see how the model's design, the network's structure, and even how we phrased our instructions (prompts) influenced interaction. What we found was interesting: Mistral and Llama3 were highly cooperative, while Phi3 tended to be more self-interested.
Interestingly, the random network of Erdős-Rényi boosted cooperation. And here is a more fascinating one: how we designed the prompts really mattered, with prompts focusing on "society benefits" leading to much higher levels of cooperation. These results give us a preliminary framework for understanding how LLMs cooperate in adaptive social networks, offering valuable insights into designing AI systems that can effectively team up with each other and with us.

\section{Related Work}

Understanding cooperation among intelligent agents in complex societal systems is a fundamental challenge in complex systems and AI research. The \textbf{structure of the network}, the way agents are connected, plays a critical role in shaping collective outcomes. Classic models, such as the \textit{Watts-Strogatz small-world network}, reveal that even modest rewiring in a regular lattice can drastically reduce path lengths, balancing local clustering with global connectivity, which influences the spread and sustainability of cooperation~\cite{watts1998collective}. Empirical and theoretical studies, including Rahwan’s work, show how network topology affects trust, information flow, and the balance between cooperation and free-riding in human groups~\cite{rahwan2019machine, crandall2018cooperating}. These insights inform the modeling of AI agents, particularly in Public Goods Games (PGGs) and related social dilemmas.

Groundbreaking research by Nicholas Christakis and his Human Nature Lab ~\cite{christakis2018social} explores hybrid human-AI networks, where AI bots—often simple “dumb AI”—are embedded within human social networks. Their studies show that strategically placed AI agents can enhance group coordination, creativity, and cooperation by leveraging network properties such as clustering and modularity.

Christakis's research adds another crucial layer to important discoveries of how humans and machines can work together by Rahwan ~\cite{rahwan2019machine, awad2018moral}. It highlights that AI's impact on society is not just about how smart its algorithms are; it also critically depends on where that AI sits within a network and how well connected it is ~\cite{wang2024llm, shirado2019network}.

Traditional AI approaches to PGGs have focused on reinforcement learning (RL) in static or simple networks ~\cite{leibo2017multi, han2024deep, Luke2005}. More recently, large language models (LLMs) such as GPT-4 and Llama-2 have shown strategic, context-aware reasoning in social dilemmas~\cite{wang2024llm, kim2024justice}. However, their applications within \textit{dynamic network settings} remain underexplored, especially in cooperative rather than purely competitive contexts. Adaptive network features like rewiring and heterogeneous connectivity—highlighted by game-theoretic studies on evolving networks~\cite{antonioni2017public, szolnoki2021rl}—are key to advancing these models. Simulation platforms such as Empirica facilitate large-scale experiments combining network dynamics and AI agents but have mainly used simple rule-based bots rather than advanced LLMs~\cite{shirado2019network}.

In this work we aim to pioneer such integration, leveraging Empirica’s flexible simulation environment and the Ollama LLM framework to explore novel cooperative dynamics. By doing so, we address important gaps in understanding how complex network topology and advanced AI reasoning combine to shape collective intelligence and social welfare.

\section{Methodology}

Using Large Language Models (LLMs) as players in a public goods game where connections are always changing, we investigate how AIs interact with each other. We consider $8$ AI agents, each starting with \$1.00, playing 16 rounds. The results represent the average across multiple runs of each experiment. Every round unfolds in two stages: in the action stage, agents decide whether to cooperate or defect, and then in the rewiring stage, agents get to reshape their social circle.
\begin{itemize}
\item \textbf{Action stage}: Deciding whether to cooperate or defect. Figure \ref{fig:action-stage} depicts how the game works.
\begin{itemize}
    \item 
Cooperation: An agent's budget decreases by \$0.05 for each connected neighbor, while each of those neighbors' budgets increases by \$0.10. An agent must defect if its current budget is insufficient to cover the cooperation cost.
\item
Defection: An agent's budget remains unchanged.
The decision-making for cooperation or defection is solely handled by different LLM "strategies". Each LLM receives information about its current budget, neighbors, and its neighbors' previous round actions to inform its decision. The specific LLM "strategy" is defined by its pre-set prompt, allowing us to explore various behavioral policies (e.g., self-interest, collective welfare).
\end{itemize}

\begin{figure*}[t]
\centering

\subcaptionbox{Snapshot Shows a Part of the Network\label{fig:act-a}}{%
    \includegraphics[width=0.32\textwidth]{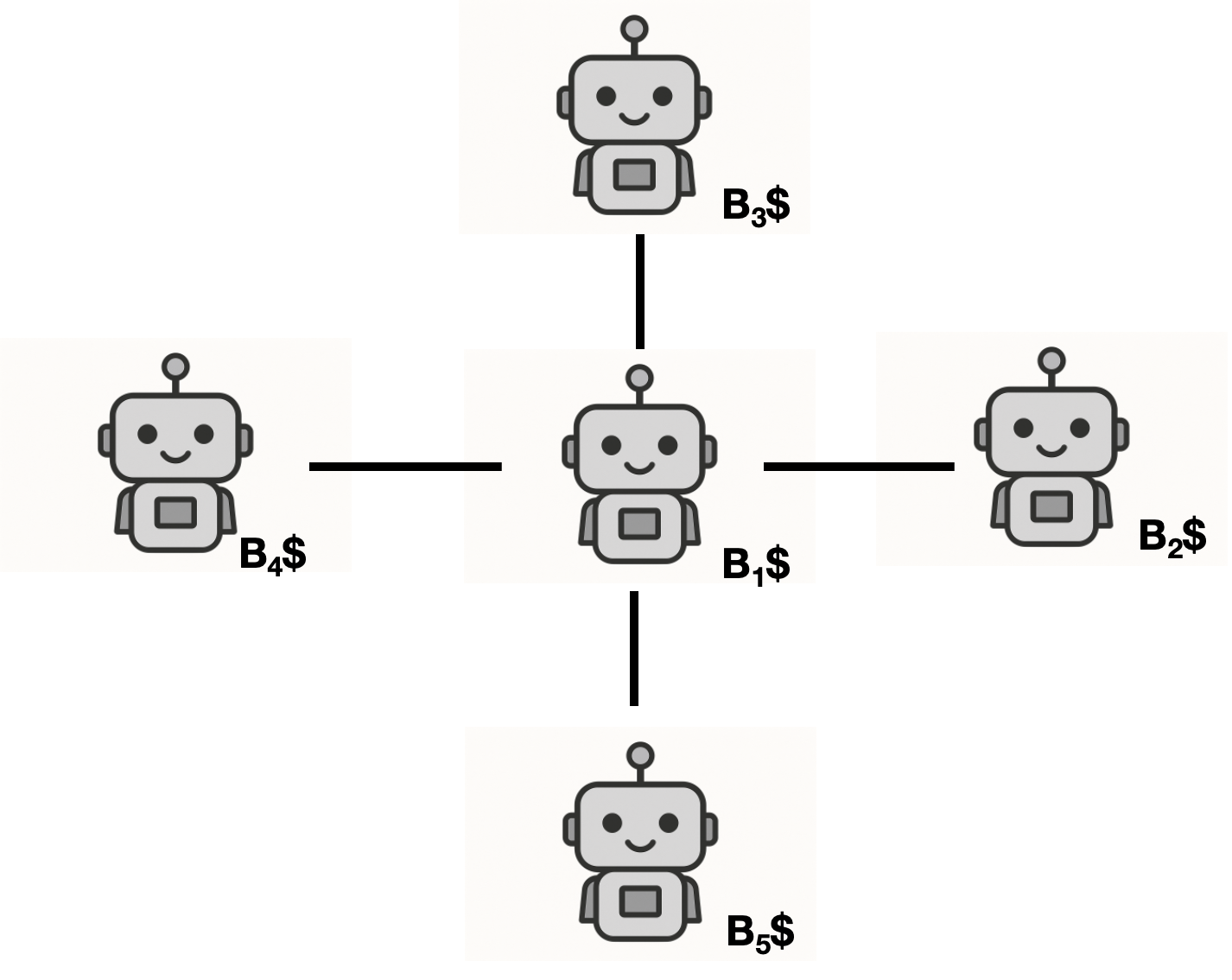}%
}
\hfill
\subcaptionbox{Central AI Chooses to Cooperate\label{fig:act-b}}{%
    \includegraphics[width=0.32\textwidth]{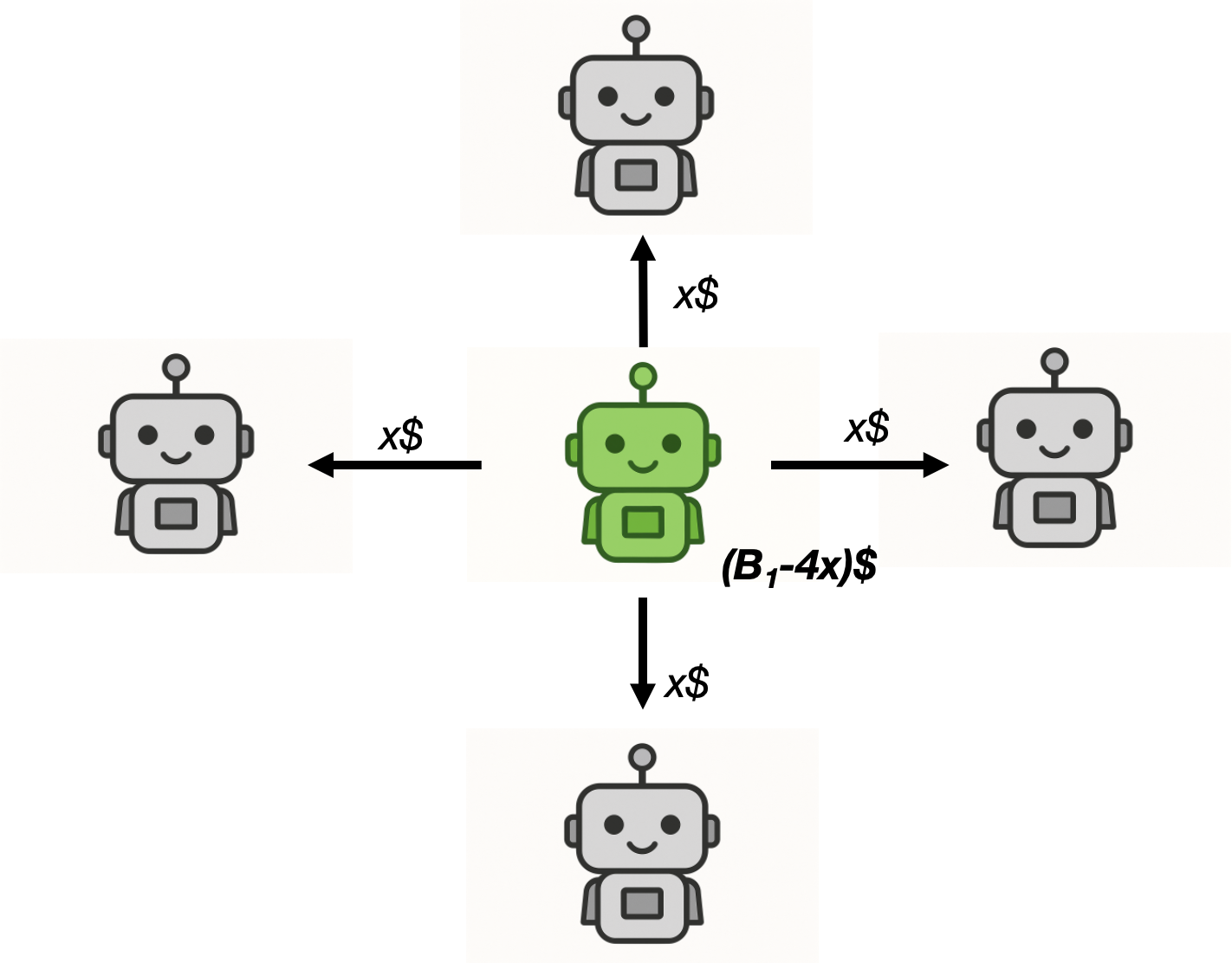}%
}
\hfill
\subcaptionbox{Three Neighbors Decide to Cooperate\label{fig:act-c}}{%
    \includegraphics[width=0.32\textwidth]{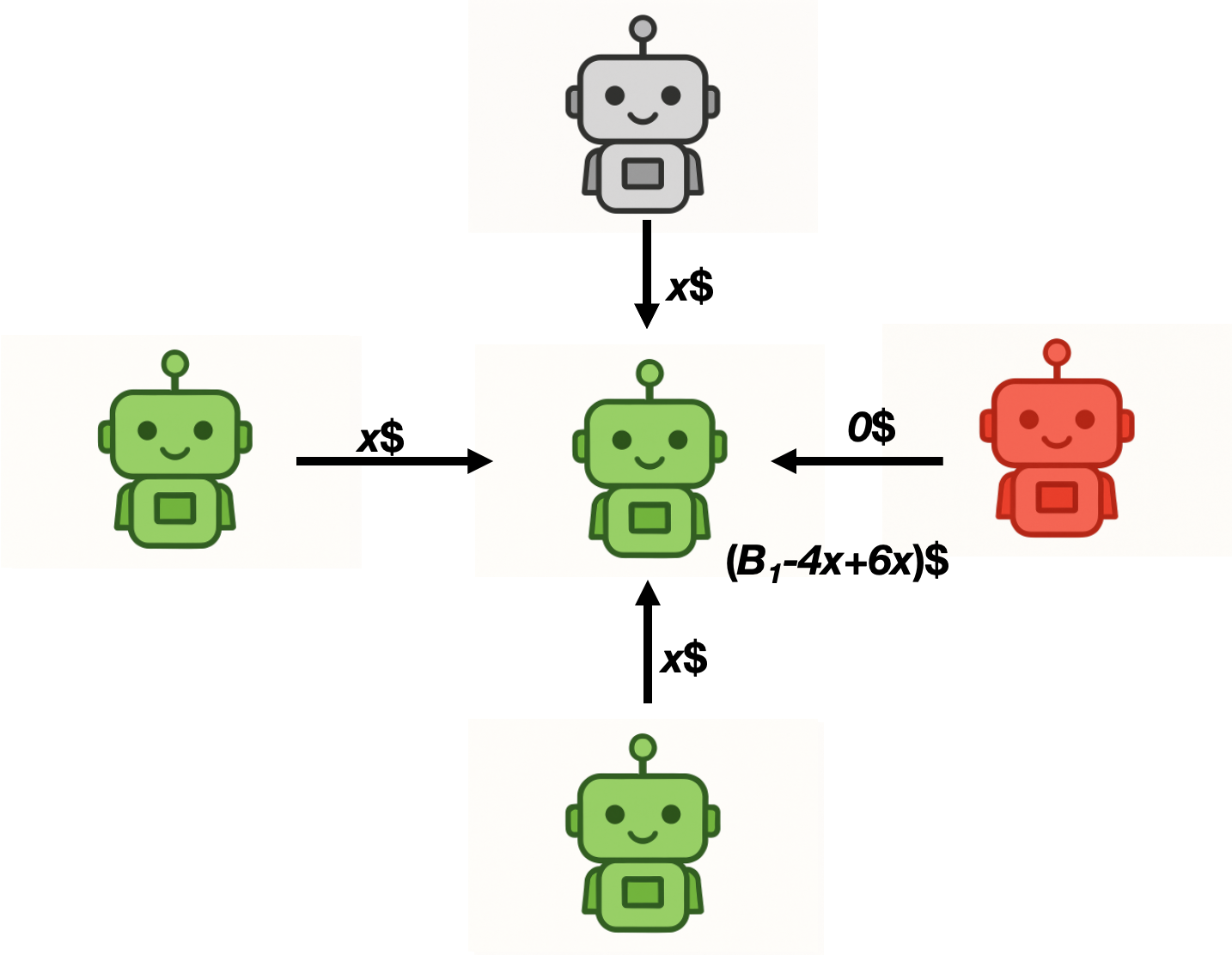}%
}

\caption{Action Stage: Deciding Whether to Cooperate or Defect. (a) This Snapshot Shows a Part of the Network Where One AI Sits at the Center, Connected to Four Neighboring AIs. (b) The Central AI Chooses to Cooperate, Contributing $x\$$ to Each of Its Four Neighbors. As a Result, Its Own Budget Drops to $(B_1-4x)\$$. (c) At the Same Time, Three Neighbors Decide to Cooperate, While One Chooses to Defect. The Defector Contributes Nothing but Still Receives $2x\$$ from Each Cooperator—Earning a Total of $6x\$$ and Ending Up with $(B_1-4x+6x)\$$.}
\label{fig:action-stage}
\end{figure*}
\begin{figure*}[t]
\centering

\subcaptionbox{Add New Neighbors?\label{fig:rew-a}}{%
    \includegraphics[width=6cm,height=4cm]{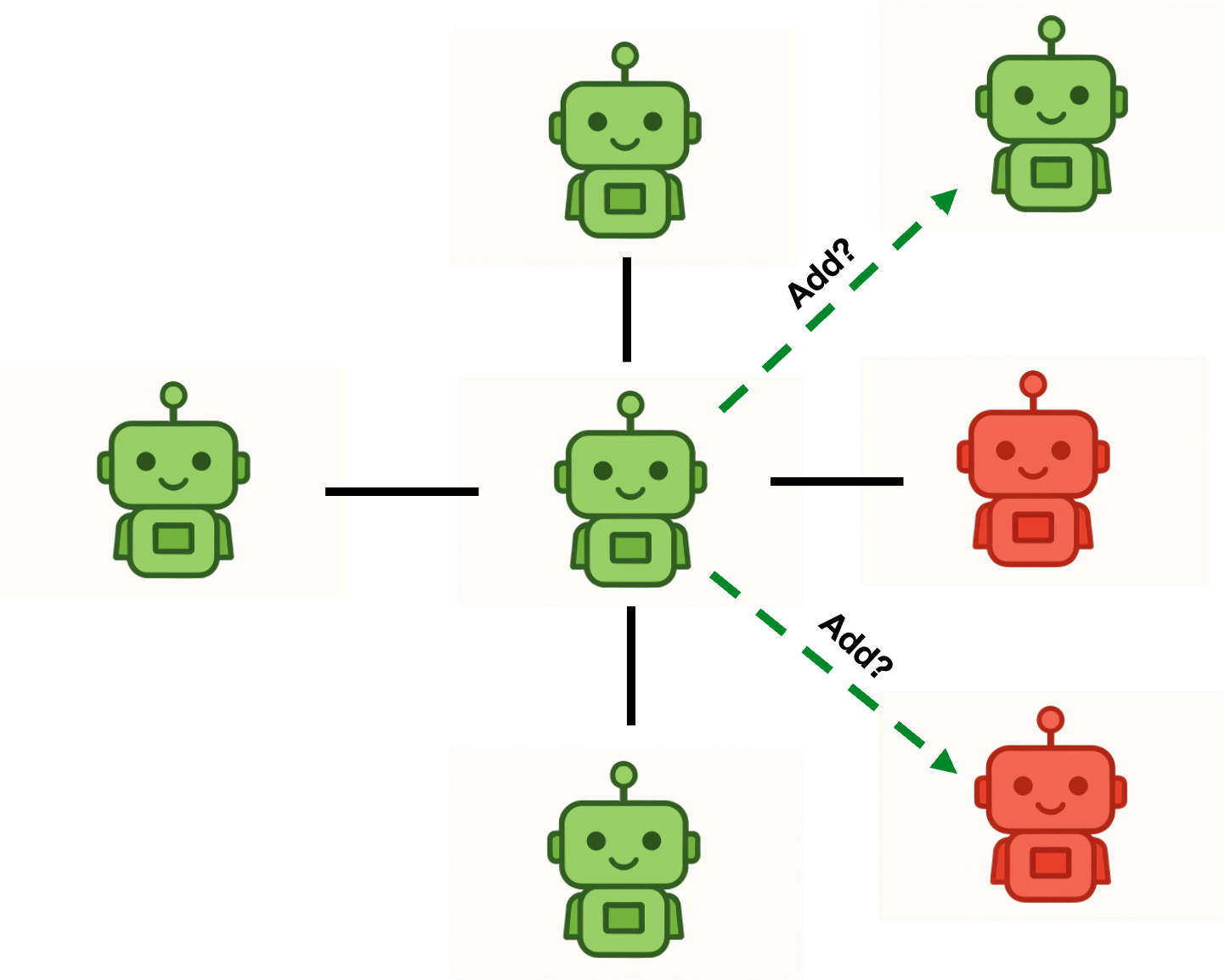}%
}
\hfill
\subcaptionbox{Remove Any Connection?\label{fig:rew-b}}{%
    \includegraphics[width=6cm,height=4cm]{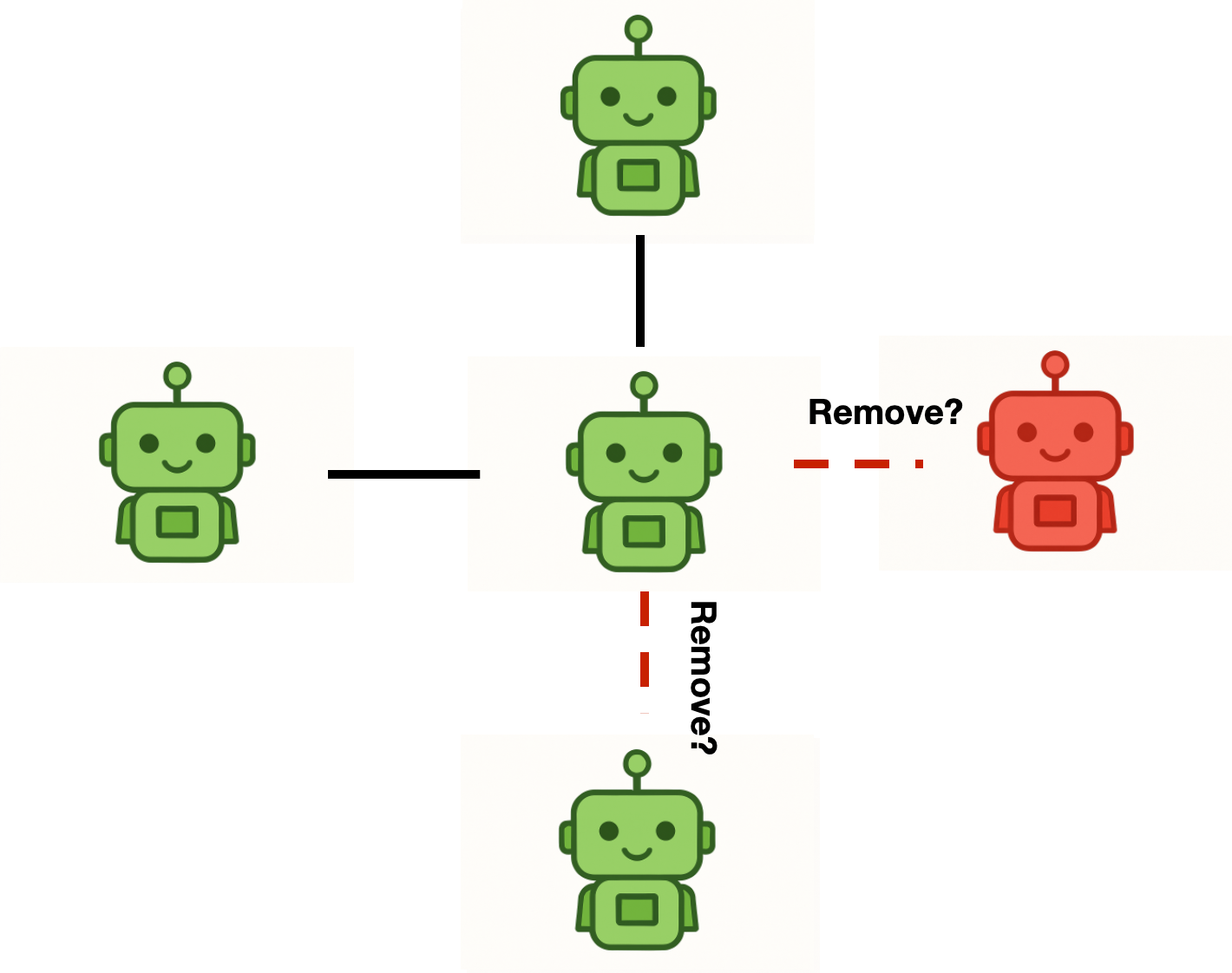}%
}

\caption{Rewiring Stage: Deciding Whether to Add or Remove the Connections.}
\label{fig:rew-stage}
\end{figure*}

\item
\textbf{Rewiring Stage}:

Following the action stage, the network dynamically evolves. Based on a rewiring parameter, we randomly select either one of an agent's neighbors (to potentially remove) or a random non-neighbor (to connect with potentially)
Agents decide to keep existing connections, drop existing connections, or form new connections. This rewiring mechanism is consistent across all experimental conditions, allowing us to isolate the effects of initial network structure and LLM strategies.

\item
\textbf{Experimental Conditions}: 
To compare agent behavior, the experiment is conducted under two primary conditions, varying the initial network structure—Watts-Strogatz (${\beta}=0.3$), Barabási-Albert ($m_0 = 3$), and Erdős-Rényi ($p=0.3$)—and the LLM agent strategies (llama3:latest, phi3:mini, mistral:latest, and gemma3:latest). The payoff structure and the rewiring rules (though not the outcomes of rewiring, which depend on agent choices) remain constant across all conditions.
\end{itemize}

This design enables us to analyze how different initial network topologies interact with diverse LLM-driven behavioral strategies to influence cooperation rates, network evolution, and overall collective welfare over time.

To understand how our agents would behave, we set up experiments with two different environments:
\begin{itemize}
\item
\textbf{All-the-Same Model}: In this setup, all eight agents used the exact same large language model. This setup served as a baseline to observe how agents behave when powered by an identical LLM.

\item
\textbf{Mixed Model Crew}: In this setup, we introduced diversity by assigning four different large language models across the eight agents—each model powering two agents. This allowed us to study how differences between LLMs impact agent behavior, coordination, and problem-solving within the system.
\end{itemize}

To assess both individual and collective motivations—at the level of single agents and the group as a whole—we gave the agents three distinct types of instructions. Before doing so, we first explained the concept of the Public Goods Game to ensure all agents had a shared understanding of the task context.:
\begin{itemize}
\item
\textbf{General Prompt}: This prompt is neutral, like, "Go ahead and act." This helped us see what their natural tendencies were without any special nudges. 

\textbf{Prompt sample}:
\begin{itemize}
    \item 

Action prompt: You play a public goods game on a dynamic network. Each round, choose to cooperate (co) by paying \$0.05 per partner so they get \$0.10 each, or defect (def) to keep your money and give partners nothing. Start with \$1.00. Partners choose simultaneously. Respond only with "co" or "def". For the first round, just choose "co" or "def". For later rounds, decide based on neighbors’ previous choices.
\item 
Remove prompt: Decide whether to "keep" or "remove" a connection based on a neighbor’s previous action. Respond with exactly one word.
\item 
Add prompt: Decide whether to "add" or "skip" a player to your network based on their previous action. Respond with exactly one word.
\end{itemize}
\item
\textbf{Self-Benefit Prompt}: Here, we specifically told them to be a bit selfish! We encouraged agents to prioritize maximizing their individual wealth, emphasizing personal gain. 
To do this, we introduce choices like "cooperate or defect to maximize your wealth," "keep or remove to maximize your wealth," and "add or skip to maximize your wealth" into the action prompt, add prompt, and remove prompt sections of the general prompt, respectively. Each choice is designed to help users think strategically about increasing their wealth.

\item
\textbf{Society-Benefit Prompt}: Finally, we encouraged them to be team players. This prompt urged them to think about the bigger picture and try to make sure everyone in the group, including themselves, ended up better off.

To encourage people to maximize total society wealth, we introduce choices like "cooperate or defect to maximize society wealth," "keep or remove to maximize society wealth," and "add or skip to maximize society wealth" into the action prompt, add prompt, and remove prompt sections of the general prompt, respectively. Each choice is designed to help users think strategically about increasing their wealth.
\end{itemize}

\section{Results}
To assess how different large language models (LLMs) behave in dynamic social dilemmas, we conducted two experiments—one using mixed LLM models and another using a single LLM model per run—across three network topologies: Watts-Strogatz (WS), Barabási-Albert (BA), and Erdős–Rényi (ER). We report findings on cooperation behavior, network evolution over time, and the effect of prompt design on model behavior. Importantly, agents were not informed about the underlying network structure, the number of rounds, or whether the interaction would be finite or infinite—forcing them to reason and adapt based solely on local interactions and observed outcomes.

\subsection{Cooperation Rate}

According to Table \ref{tab:mixed_model}, in the mixed-model experiment, Mistral demonstrated near-perfect cooperation across all networks ($0.94$–$1.00$), while Phi3 showed the lowest rates ($0.28$–$0.37$). In the single-model experiment (Table \ref{tab:single_model}), Mistral maintained complete cooperation on all networks, and Llama3 was nearly perfect as well—only slightly lower on the ER network ($0.86$). In contrast, Gemma3 and Phi3 showed much lower cooperation levels. Notably, in both experiments, all models (except Llama3) exhibited the highest cooperation rates on the ER network, likely due to its randomized structure enabling broader agent interaction \cite{Holme2012}.

\begin{table*}[!htbp, width=0.48\textwidth]
    \centering
    \begin{minipage}[t]{0.48\textwidth}
        \centering
        \caption{Cooperation Rates per Model Across Different Network Types in the Mixed-Model Experiment}
        \label{tab:mixed_model}
        \begin{tabular}{lccc}
            \toprule
            \textbf{Model} & \textbf{WS} & \textbf{BA} & \textbf{ER} \\
            \midrule
            Llama3 & 0.97 & 0.97 & 0.87 \\
            Mistral & 0.94 & 1.00 & 1.00 \\
            Gemma3 & 0.44 & 0.53 & 0.81 \\
            Phi3 & 0.28 & 0.31 & 0.37 \\
            \midrule
            Overall & 0.66 & 0.70 & 0.76 \\
            \bottomrule
        \end{tabular}
    \end{minipage}%
    \hfill
    \begin{minipage}[t]{0.48\textwidth}
        \centering
        \caption{Cooperation Rates per Model Across Different Network Types in the Single-Model Experiment}
        \label{tab:single_model}
        \begin{tabular}{lccc}
            \toprule
            \textbf{Model} & \textbf{WS} & \textbf{BA} & \textbf{ER} \\
            \midrule
            Llama3 & 1.00 & 1.00 & 0.86 \\
            Mistral & 1.00 & 1.00 & 1.00 \\
            Gemma3 & 0.19 & 0.47 & 0.76 \\
            Phi3 & 0.32 & 0.40 & 0.43 \\
            \bottomrule
        \end{tabular}
    \end{minipage}

\end{table*}

\subsection{Network Dynamics and Prompt Impacts}

\begin{figure*}[t!]
\centering

\subcaptionbox{Degree\label{fig:degree_multi}}{%
    \includegraphics[width=7cm,height=4cm]{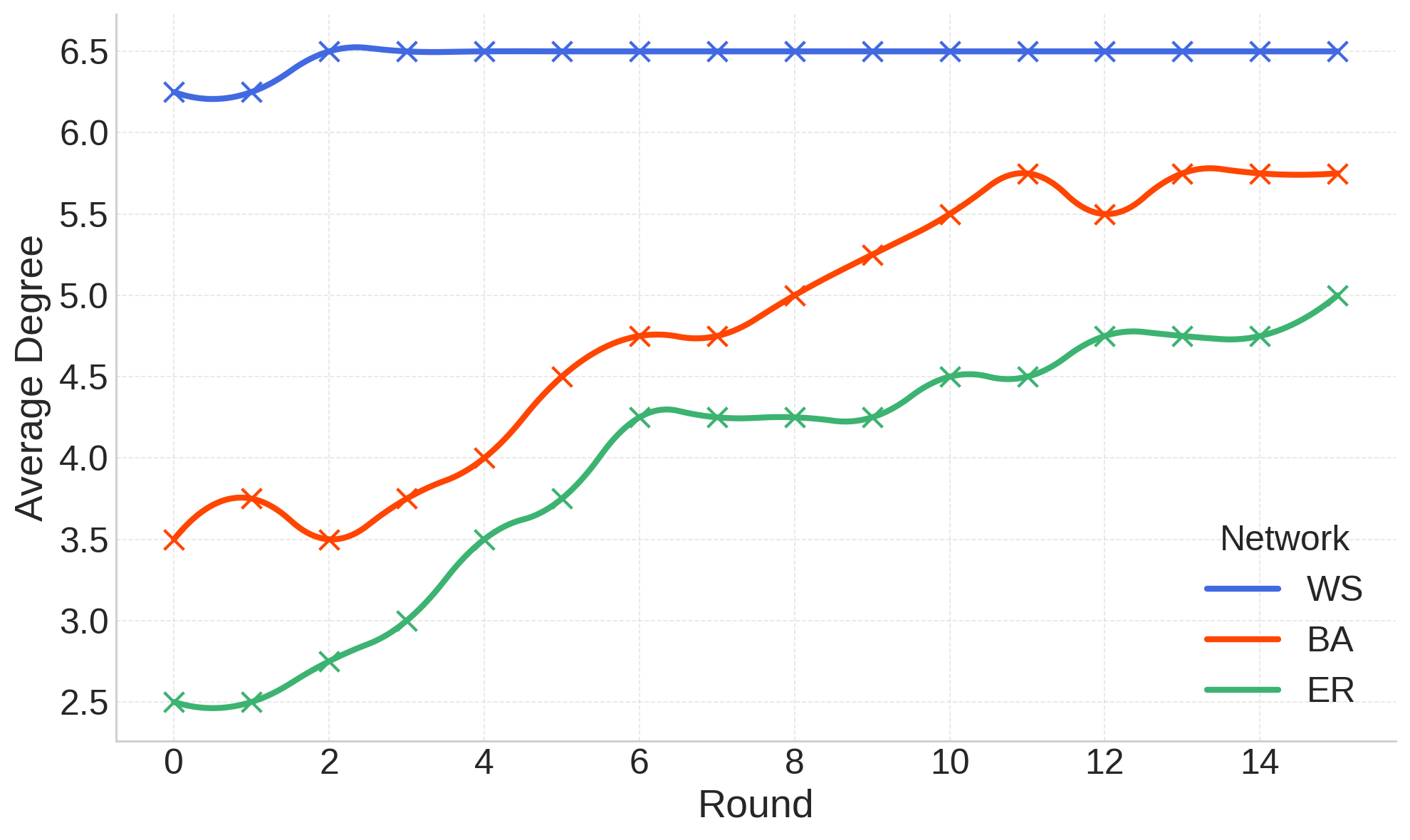}%
}
\hfill
\subcaptionbox{Clustering\label{fig:coefficient_multi}}{%
    \includegraphics[width=7cm,height=4cm]{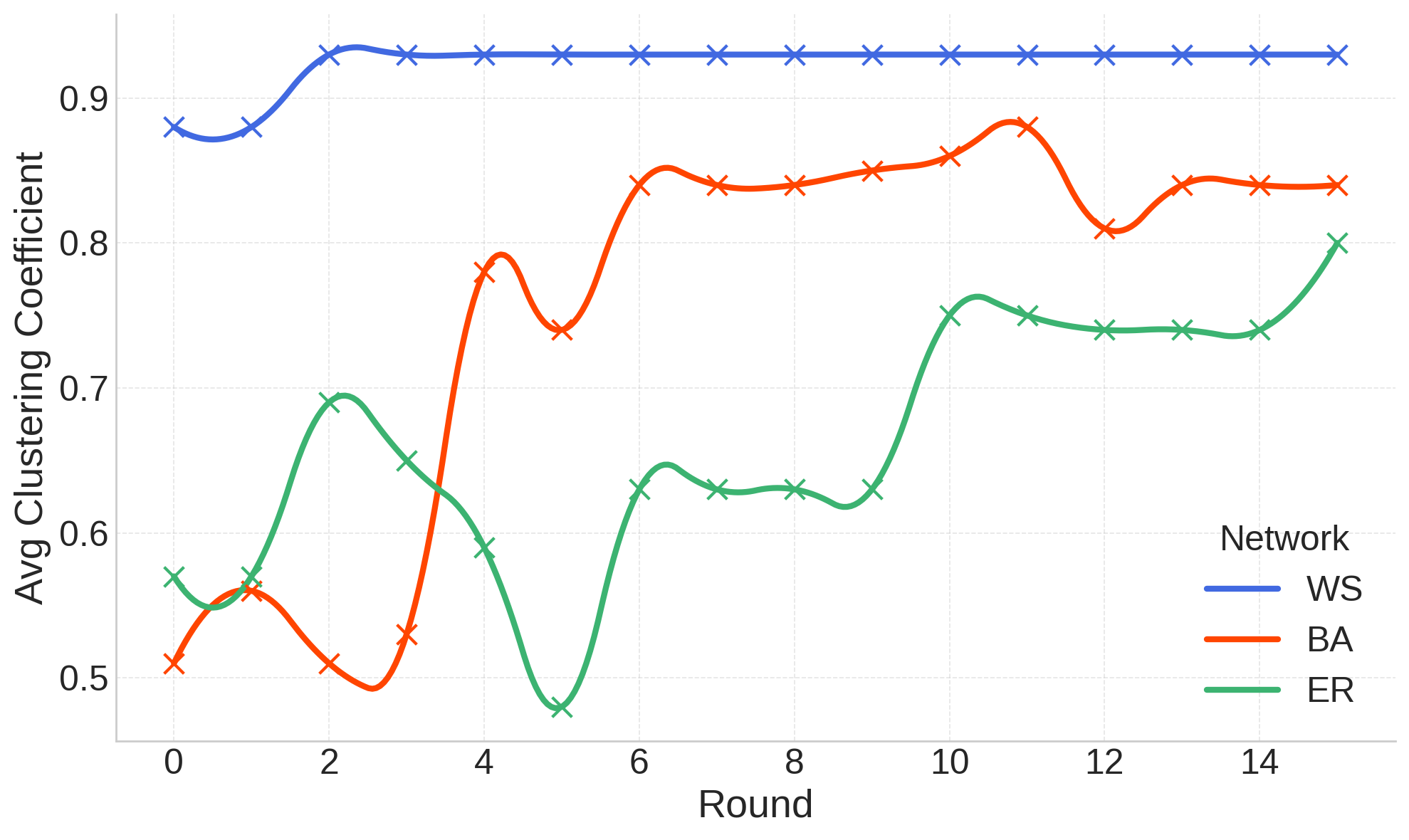}%
}

\caption{Mixed-Model Setup: (a) Degree Evolution over Rounds Across Different Network Structures. (b) Clustering Coefficient Evolution over Rounds Across Different Network Structures.}
\label{fig:results}
\end{figure*}

Our analysis of the mixed-model experiment revealed distinct evolutionary paths for each network type.
In the mixed-model experiment, the average degree in the WS network started at 6.25, grew slightly to 6.5 by the third round, and then remained stable. The clustering coefficient increased from 0.88 to 0.94, reflecting the typical high clustering found in WS networks. On the other hand, in the Barabási-Albert (BA) network, the average degree increased from 3.5 to 5.75 by the final round. The steady increase in clustering reveals that the rewiring mechanism enhanced overall connectivity. The ER network exhibited the most significant change, with the average degree rising from 2.5 to 5 and the clustering coefficient from 0.57 to 0.80—consistent with the properties of random graphs (Figure \ref{fig:degree_multi}, \ref{fig:coefficient_multi}).

\begin{figure}[t!]
    \centering
    \includegraphics[width=14cm,height=6cm]{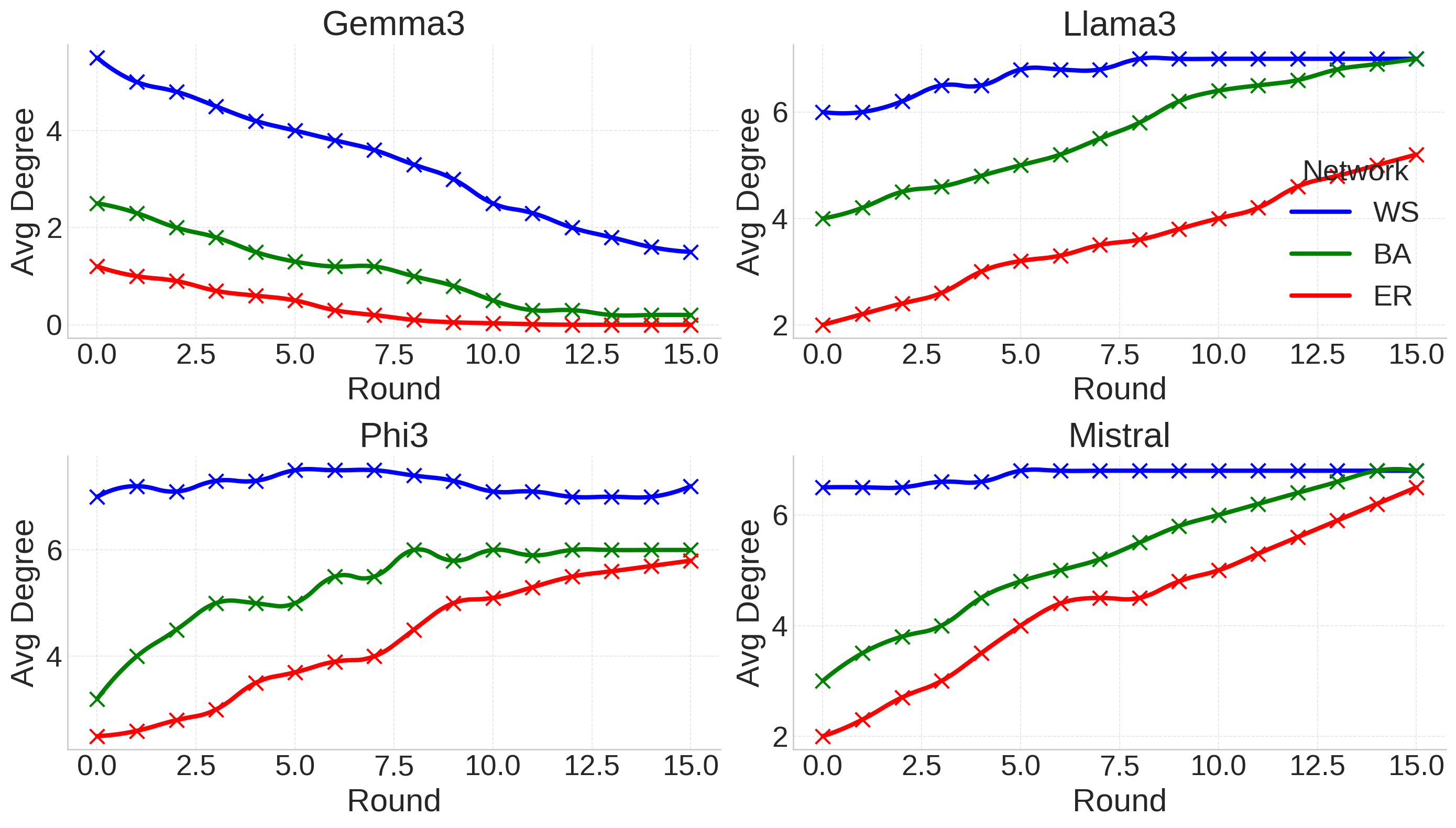}
    \caption{Degree Evolution over Rounds in the Single-Model Experiment}
    \label{fig:degree_single}
\end{figure}

\begin{figure}[t!]
    \centering
    \includegraphics[width=14cm,height=6cm]{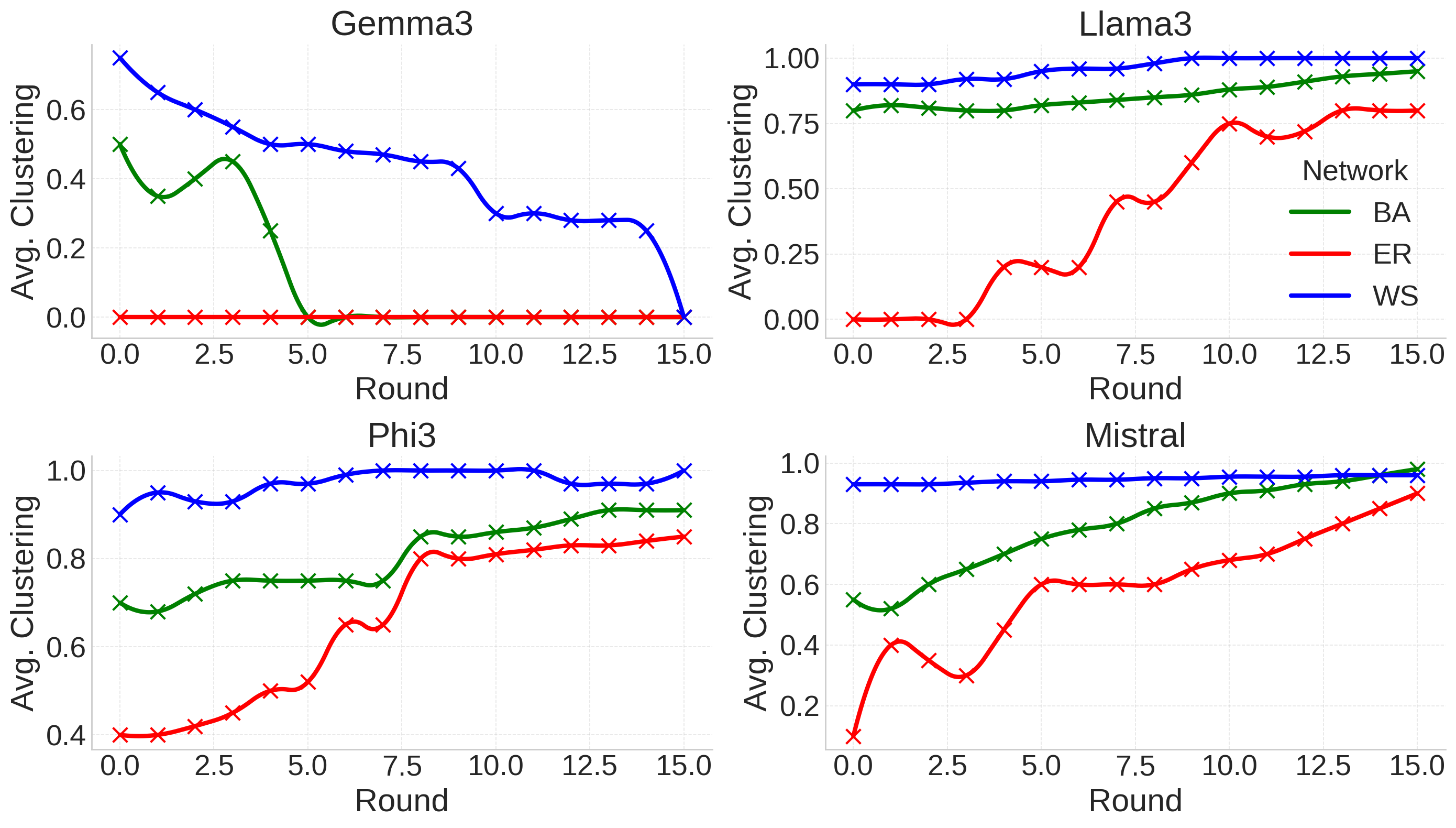}

    \caption{Clustering Coefficient Evolution in the Single-Model Experiment}
    \label{fig:clustering_single}
\end{figure}

In the single-model experiment, network properties varied more dramatically. Cooperative models (Mistral and Llama3) tended to increase both mean degree and clustering, while models with low cooperation (like Gemma3) caught declines in both (Figure \ref{fig:degree_single}, \ref{fig:clustering_single}).
These collective findings suggest that cooperative behavior among agents not only fosters stronger social ties but also drives the emergence of more cohesive and well-connected network structures, while defective behavior leads to fragmentation and reduced connectivity.

Our study illustrates that society-benefit prompts generated the highest cooperation rates with 0.93 and a mean score of 3.59, followed by self-benefit prompts (0.89, 3.40) and general prompts (0.65, 2.86). This result highlights the sensitivity of LLMs to prompt design.
	
\section{Discussion and Conclusion}
Our findings demonstrate the impact of network topology, LLM model, and prompt design on the behavior of agents and their level of cooperation in LLM-based methods. The random structure of the ER network fosters interactions and leads to cooperation. Mistral and Llama3 present themselves as reliable collaborators, while Phi3 exhibits defective behaviors, and Gemma3 is a context-dependent model. Moreover, the highest cooperation rate was achieved by the society-benefit prompt, which aligns with collective goals. These results suggest the effectiveness of lightweight LLMs for modeling adaptive social behaviors and interactions in intelligent systems.
	
This work presents a scalable framework for intelligent systems by using lightweight LLMs and PGGs. We highlight the importance of model architecture, prompt design, and network topology in changing the cooperative behavior of agents. Future studies can explore more robust LLMs, human-AI interactions, and richer prompts to enhance cooperative AI systems.
	
% ...
	
\section*{Declaration on Generative AI}
During the preparation of this work, the author(s) used GPT-4 and Grammarly in order to: Grammar and spelling check. After using these tool(s)/service(s), the author(s) reviewed and edited the content as needed and take(s) full responsibility for the publication’s content.


\bibliography{main}


\end{document}